\documentclass[preprint,12pt]{elsarticle}
\usepackage{amssymb}
\usepackage{amsmath}
\usepackage{graphicx}
\usepackage{dcolumn}
\usepackage{bm}
\usepackage{xcolor}
\usepackage{caption}
\usepackage{float}
\usepackage{booktabs}\usepackage{multicol}
\journal{Brazilian Journal of Physics}
\begin{document}
\begin{frontmatter}
\title{Re-investigation of neutron capture by $^{84}$Kr and $^{86}$Kr in the $s$-process nucleosynthesis.}
\author{{Abdul Kabir$^{1}$, Zain Ul Abideen$^{1}$,  and  Jameel-Un Nabi$^{2}$.}}
\address{$^{1}${Space and Astrophysics Research Lab, National Centre of GIS and Space Applications, Department of Space Science, Institute of Space Technology, Islamabad 44000, Pakistan.}}
\address{$^{2}${University of Wah, Quaid Avenue, Wah Cantt 47040, Punjab, Pakistan.}}

\begin{abstract}
The thermonuclear reaction rates and Maxwellian-averaged cross-sections (MACS) for the $^{84}$Kr(n,$\gamma$)$^{85}$Kr and $^{86}$Kr(n,$\gamma$)$^{87}$Kr processes were examined using the statistical model code Talys v1.96. The effects of nuclear level {densities (NLDs) on Maxwellian-averaged cross-sections} and neutron capture rates are examined both quantitatively and qualitatively. The present {Talys-based } MACS and radiative capture rates for $^{84}$Kr(n,$\gamma$)$^{85}$Kr and $^{86}$Kr(n,$\gamma$)$^{87}$Kr processes {agree} well with the earlier reported findings. The statistical model's nuclear properties (level density and gamma-ray strength) were fine-tuned to reproduce the existing experimental nuclear data.  
\end{abstract}
\begin{keyword}
Cross-section, AGB stars, MACS, Talys v1.96, Nuclear Level Density, Nuclear Rates.
\end{keyword}
\end{frontmatter}
\section{Introduction:} 
Nuclei that are more massive than iron (Fe) are mostly formed by neutron capture reactions, which are called after their widely distinct time scales as the rapid neutron capture process ($r$-process) and the slow neutron capture process ($s$-process). The rapid neutron capture process takes place in a neutron-rich very hot stellar environment ($\rm T_{9}$ $>$ 1). The slow neutron capture process operates at a lower temperature and lower neutron density $(\rho_n)$. Heavy elements are synthesized in stars via the $r$-process and $s$-process, {and} these two distinct mechanisms have distinct characteristics.  As mentioned above, they both differ in their time scales and  $\rho_n$~\cite{Burbidge}. The $s$-process involves a series of successive captures and decays, occurring in the region close to the valley of stability, and it has received substantial experimental and theoretical attention \cite{Kappeler,Pignatari,Kappeler1}.  Because of breakthroughs in astronomical observations and stellar modeling, accurate assessments of the $s$-process have become incredibly influential~\cite{Lugaro}.  The $s$-process has grown from a simple explanation of the abundance distribution of elements in the solar system to a deeper understanding that incorporates broad stellar and galactic dynamics. Because of the developments, the $s$-process is now a powerful tool for understanding the evolution of red giant stars~\cite{Busso}. 

The reader's attention is drawn here to the noticeable cross-sections of the krypton (Kr) nuclei. Significant progress has been made in the examination of Kr, both from a nuclear astrophysical perspective \cite{Tessler, Mutti}. Still, there is much to learn about this important element, and further research into its properties and behavior is necessary to deepen our understanding of nuclear and astrophysical processes. In multiple contexts,  the Kr isotopes are of particular significance. They inhabit a nuclide region that is susceptible to a variety of $s$-processes, for example, the coined weak (56$ \leq$$A$$\leq$ 90) and the main ($A$$\ge$90), ascribed to the emergence of huge and low-mass stars, respectively. Usually, stellar neutron capture rates for $s$-process investigations are based on a  {typically} consistent thermal energy of $kT$=30 keV, which corresponds to $T_9$=0.35. The cross-sections of nuclei at magic neutron numbers are described by a few resonances in the energy range $kT$=30 keV, which may lead to an unusual energy dependence of the Maxwellian average. High precision is required because these isotopes act as barriers in the neutron capture pathway of the $s$-process due to their relatively modest cross-section values. 

Many researchers in the past  {have analyzed} the impact of the $s$-process on Kr isotope abundances, precise neutron capture cross-sections for both stable and unstable nuclides in s-process scenarios, as described below. Still, the interest in the topic has continued  {unabated to date}. Walter \textit{et al.} \cite{Walter} analyzed the neutron capture cross-sections of the stable Kr isotopes in the energy range (4--250)~keV by employing a $\rm C_6D_6$-detector device and the time-of-flight approach.  The investigations were augmented by statistical model estimates of all Kr isotopes within the 78$<$A$<$86 ranges, to produce an acceptable cross-section for the unstable nuclei of Kr. For all important parameters, the estimates were based on local systematics, and the findings were found to have (20-25)\% uncertainty. Their Maxwellian   cross-sections at $kT$= 30 keV were determined to be 36$\pm$4.5~mb.  Tomyo \textit{et al.} \cite{Tomyo} analyed the discrete $\gamma$-ray emitted promptly by the $^{84}$Kr(n,$\gamma$)$^{85}$Kr. They have accurately {determined} the neutron capture cross-section by a NaI(T1) spectrometer between (10--80)~keV. Their resultant {cross-section agreed }with earlier  {existing} data. Kappeler \cite{Kappeler9}  investigated the neutron capture cross-section in the domain of $s$-process of nucleosynthesis.
They measured the neutron capture cross-section at $kT$=25 keV for $^{84}$Kr(n,$\gamma$)$^{85}$Kr and $^{86}$Kr(n,$\gamma$)$^{87}$Kr reactions. Their analyzed cross-section at $kT$=30 keV were  16.7$\pm$1.2 and 3.5$\pm$0.3~mb, respectively. {Tessler \textit{et al.} \cite{Tessler}} analyzed the neutron capture cross-section for a number of Kr nuclei subjected to measurement by the activation framework. Their measurements were performed in the intense $\approx$ 40 keV quasi-Maxwellian neutron field of the SARAF-LiLiT facility. From the experimental data, they have extracted the Maxwellian {averaged} cross-section. Their measured value of MACS at $kT$= 30 keV is 34.7(28)~mb. Mutti \textit{et al.}~\cite{Mutti} evaluated a number of neutron capture cross-section measurements on Kr nuclei that were carried out at the Geel Electron Linear Accelerator (GELINA). They measured the total neutron capture cross-section by $^{82}$Kr, $^{84}$Kr, and $^{86}$Kr nuclei  {around} E$_n$=400 keV. Their primary goal was to produce reputable nuclear data for $s$-process nucleosynthesis simulations. 

In our present investigations, both the $^{84}$Kr(n,$\gamma$)$^{85}$Kr and $^{86}$Kr(n,$\gamma$)$^{87}$Kr reactions have been analyzed within the framework of statistical model code Talys v1.96~\cite{TALYS1.96}.  The Talys v1.96  {code} is frequently used to analyze nuclear reaction data, especially for reactions caused by nucleon capture. The model's prediction was compared to the measured data. {The Talys v1.96 is based on the Hauser-Feshbach Theory {\cite{hf}}. The Optical Model Potential (OMP), Nuclear Level Density (NLD), and the Radiative Strength Function (RSF) are the main inputs of the theory. For reactions involving low-energy neutrons, the effects of OMP can be ignored in favor of the NLD and RSF \cite{ompignore}. The optical model employed in this study is the local OMP by Koning and Delaroche \cite{localomp}}. {In the present investigation,} the MACS of the above-noted reactions have been computed up to $E_n$= 0.1 MeV. The abundances of the $s$- and $r$-processes may be reproduced via the sensitivity simulations. However, one of the primary factors of uncertainty involves, the accurate assessment of nuclear parameters such as the neutron capture rates at which the nuclei participate in $s$- or $r$- processes.

\section{Theoretical Framework}
{The Talys v1.96 code used for the simulation of nuclear reactions, includes several state of the art nuclear models to cover almost all key reaction mechanisms encountered for light particle-induced nuclear reactions. It provides an extensive range of reaction channels and observables. {The possible incident particles in the $E_i$=(0.001–200)~MeV, the target nuclides from $A$=12 and onwards \cite{TALYS1.96}}. The output of the nuclear reaction includes fractional and total cross-sections, angular distributions, energy spectra, double-differential spectra, MACS, and capture rates.  Radiative capture is important in the context of nuclear astrophysics in which a projectile {fuses} with the target nucleus and {emits $\gamma$- ray} \cite{Kabir1,Kabir2,Kabir3,Kabir4}. The nuclear cross-section is an important factor in the calculation of radiative capture rates. The Maxwellian averaged cross-section is used when the energies of the projectiles follow a Maxwellian distribution, like in the stellar environment. MACS is an average of the cross-section over a range of energies, weighted by the Maxwell-Boltzmann distribution.}
\begin{align}
	{\langle \sigma \rangle (kT) = \frac{2}{\sqrt{\pi}(kT)^2}\int_0^\infty E \sigma (E) exp(\frac{-E}{kT}) dE.}
\end{align}
{where  $k$, is the Boltzmann constant, $T$ is the temperature, $\sigma(E)$ is the capture cross-section and $E$ is the incident energy of the projectile.}

{In statistical models for predicting nuclear reactions, level densities are needed at excitation energies where discrete level information is not available or {is} incomplete. Together with the optical model potential, a correct level density is perhaps the most crucial ingredient for a reliable theoretical analysis of cross-sections, angular distributions, and other nuclear  quantities. There are six different level density models, among them three phenomenological and the {rest microscopic}, are consistently parameterized. For each of the phenomenological models, the Constant Temperature Model (CTM), the Back-shifted Fermi  gas model (BSFM), and the Generalized Superfluid model (GSM), a version without explicit collective enhancement is considered. In the CTM, the excitation energy range is divided into low energy regions i.e.,  {from 0 keV up to the matching energy ($E_{M}$) and high energy above the $E_{M}$ where the fermi-gas model (FGM) applies}. Accordingly, the constant temperature part of the total level density reads as.} 
\begin{align}
{\rho_{T}^{tot}(E_x) = \frac{1}{T}exp(\frac{E_x-E_o}{T})}
\end{align}
{where $T$ and $E_o$ serve as adjustable parameters in the constant temperature expression.  The BSFM is used for the whole energy range by treating the pairing energy as an adjustable parameter.}
\begin{align}
{\rho_{F}^{tot}(E_x) = \frac{1}{\sqrt{2\pi}\sigma}\frac{\sqrt{\pi}}{12}\frac{exp(2\sqrt{aU})}{a^{1/4}U^{5/4}}}
\end{align}
{where $\sigma$ is the spin cut-off parameter, which represents the width of the angular momentum distribution, $U$ is the  effective excitation energy and $a$ is the level density parameter defined below.} 
\begin{align}
{a = \tilde{a}(1+\delta W \frac{1-exp(-\gamma U)}{U})}
\end{align}
{where $\tilde{a}$ is the asymptotic level density without any shell effects. $\delta W$ gives the shell correction energy, and the damping parameter $\gamma$ determines how rapidly $a$ approaches $\tilde{a}$. One should note that for the best fitting one can readjust $a$ to achieve the {desired value of cross-section and nuclear reaction rates}. For further information, one can see~\cite{koni}. The GSM is similar to CTM in the way that it also divides the energy range into low and high energies. The high energy range is described by BSFM as  mentioned before.  {It} is characterized by a phase transition from a superfluid behavior at low energies, where the pairing correlations strongly influence the level density.} 
\begin{align}
{	\rho^{tot}(E_x) = \frac{1}{\sqrt{2\pi}\sigma}\frac{e^S}{\sqrt{D}}}
\end{align}
{where $S$ is the entropy and $D$ is the determinant related to the saddle-point approximation.
 {Apart from the above mentioned phenomenological models, microscopic level density models are also available in {Talys}. They include Skyrme-Hartree-Fock-Bogoluybov NLD \cite{shfb}, Gogny-Hartree-Fock-Bogoluybov NLD \cite{ghfb}, and the Temperature-dependent Gogny-Hartree-Fock-Bogoluybov NLD \cite{tdghfb}.}}
The RSF plays a crucial role in estimating cross-sections and reaction rates, particularly in processes involving the emission of gamma rays. From the different  {RSFs} included in Talys, the Temperature-dependent Relativistic Mean Field (RMF) model was employed for all calculations in our present investigations. See the reference \cite{rmf} for more information.

\section{Results and Discussion}
\begin{figure}[h!]\centering
	{\includegraphics[width=1.2\textwidth]{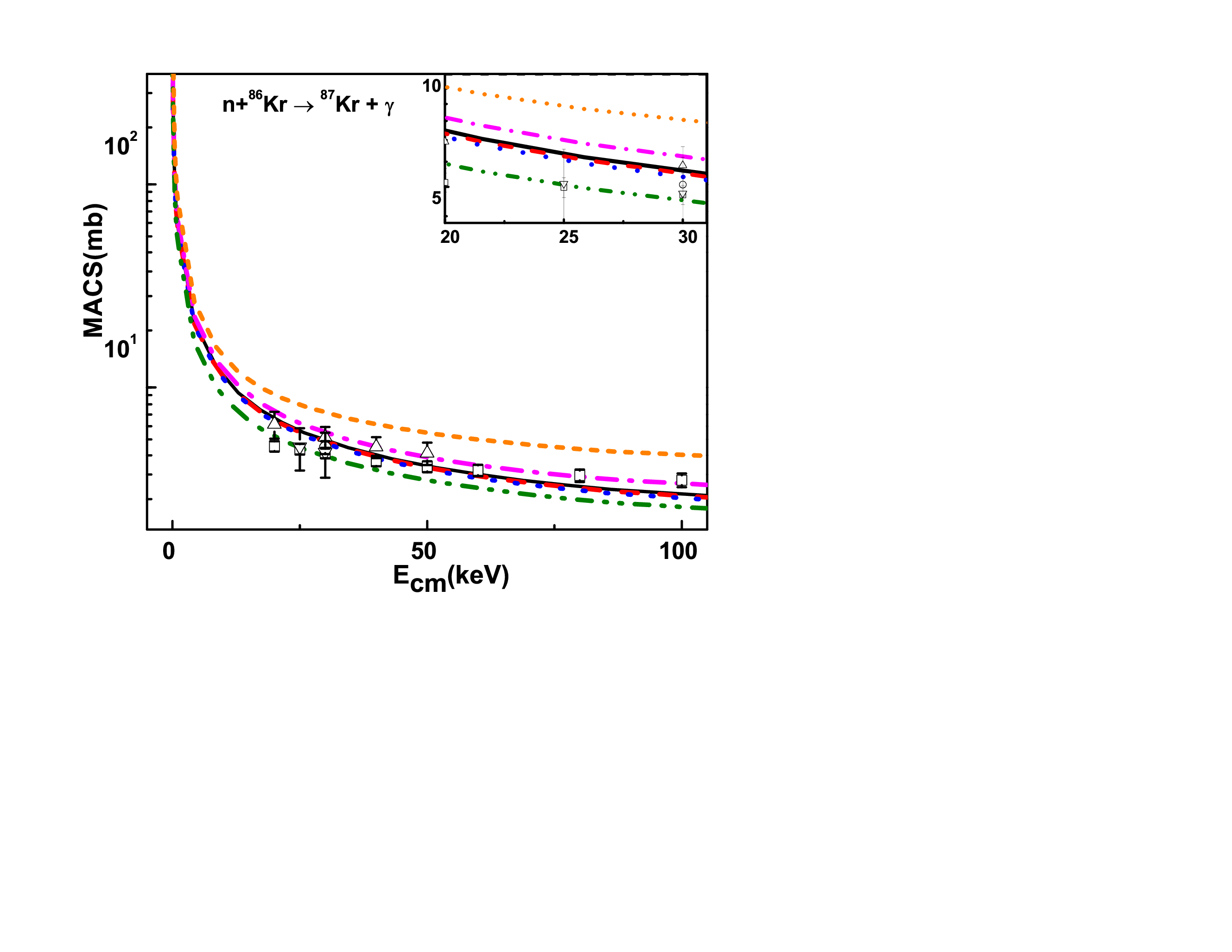}}
	\vspace*{-45mm}
	\caption{The total MACS for $^{86}$Kr(n,$\gamma$)$^{87}$Kr along with the experimental data  $\bigtriangledown$ \cite{Walter}, $\circ$ {\cite{Tessler}}, $\square$ {\cite{Mutti}},  and $\nabla$ \cite{compiled}. The solid line (black) represent the CTM, the dasehes (red) represent the BSFM, the dotted (blue) represents the GSM, the dashes dotted (magenta) represents the SHFB, the dashes dotted dotted (olive) represents the GHFB and the short dashes (orange) represents the TGHFB based computed MACS.}
	\label{fig:2}    
\end{figure}

 In the case of neutrons under stellar conditions that characterize $s$-process nucleosynthesis, a thermal equilibrium is established and one can obtain the stellar reaction rate per particle pair by averaging the measured cross-section over a Maxwell-Boltzmann distribution. The investigation of the neutron flow around the neutron magic number or near magic number $N$=50 offers the possibility of making estimates of the $s$-process neutron density and temperature. To investigate the impact of $^{84}$Kr(n,$\gamma$)$^{85}$Kr and $^{86}$Kr(n,$\gamma$)$^{87}$Kr reactions in the $s$-process of nucleosynthesis, we have computed the MACS and neutron capture rates via different nuclear level density models within the framework of Talys v1.96.
Both Kr isotopes under present investigations have almost magic neutron configurations, that result in their small (n,$\gamma$) cross-sections and fairly large $s$-abundances.  The MACS of $^{84}$Kr(n,$\gamma$)$^{85}$Kr and $^{86}$Kr(n,$\gamma$)$^{87}$Kr were computed via temperature-dependent relativistic mean field model as {RSF} and all six NLD models including the constant temperature model (CTM), back-shift fermi gas model (BSF), generalized superfluid model (GSF), Skyrme-Hartree-Fock-Bogoluybov (SHFB), Gogny-Hartree-Fock-Bogoluybov (GHFB), and Teperature dependent Gogny-Hartree-Fock-Bogoluybov (TGHFB) available in Talys v1.96. The neutron's energy was taken within (0.01-100)~keV.
\begin{figure}[h!]\centering
{\includegraphics[width=1.2\textwidth]{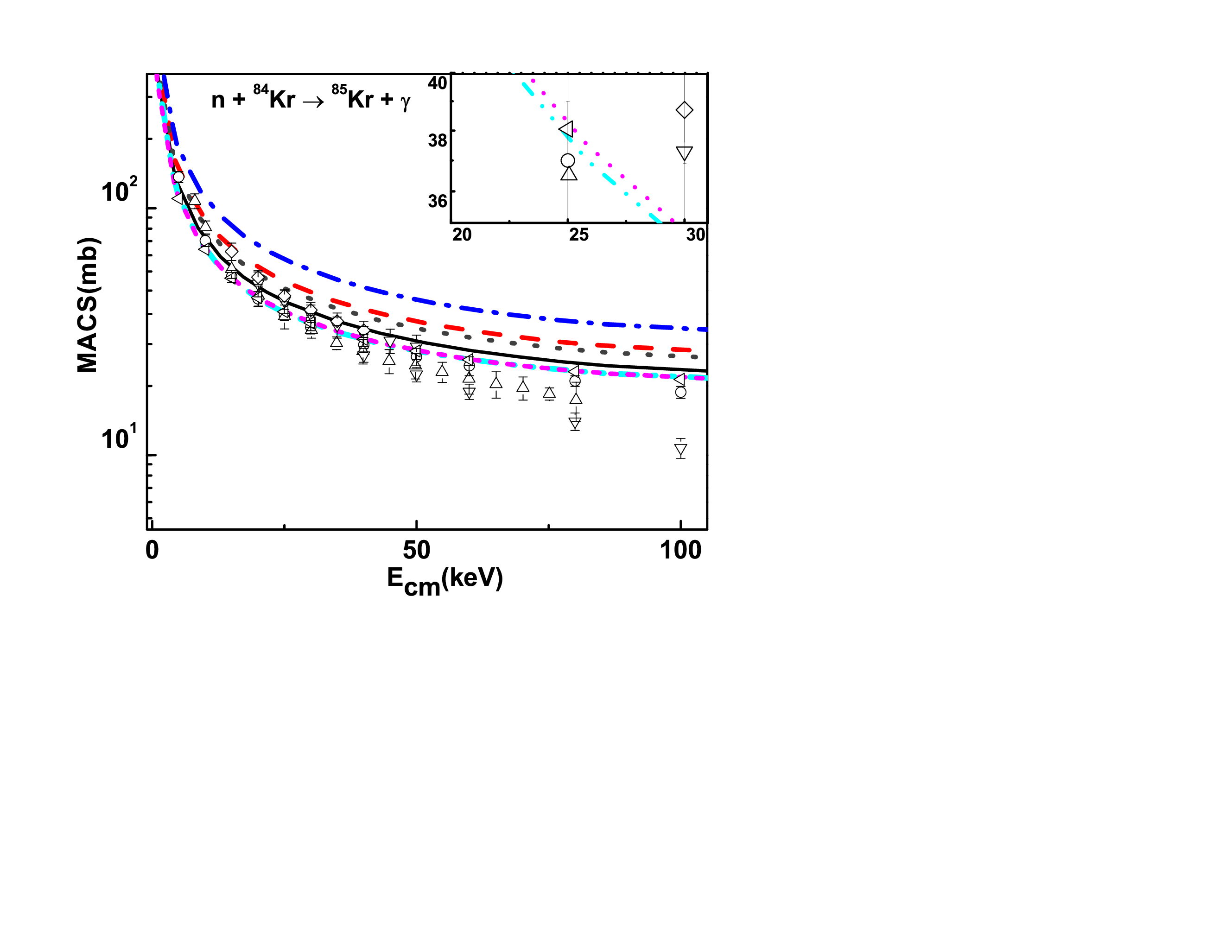}}
\vspace*{-53mm}
\caption{The total MACS for $^{84}$Kr(n,$\gamma$)$^{85}$Kr along with the experimental data $\triangleleft$ \cite{Lugaro}, $\bigtriangledown$ \cite{Walter}, $\bigtriangleup$ \cite{Tomyo}, $\square$  \cite{Tessler}, and $\diamond$ \cite{Mutti}. The solid line (black) represent the CTM, the dashes (red) represents the BSFM, the dotted (dark gray) represent the GSM, the dash dotted (blue) represent the SHFB, the dash dotted dotted (cyan) represent the GHFB and the short dasehes  (magenta) represent the TGHFB based computed MACS.}
\label{fig:1}    
\end{figure}
{$^{84}$Kr isotopes have $N$=48 neutron configurations {and} significant $s$-abundances. We aim to check the MACS at typical $s$-process temperature and the astrophysical reaction rates of the $^{84}$Kr(n,$\gamma$)$^{85}$Kr process.  The results based on the above-mentioned sets of NLDs  along with the measured data for the $^{84}$Kr(n,$\gamma$)$^{85}$Kr radiative capture process are depicted in Fig.~(\ref{fig:1}).  It can be seen that the present computed results of MACS based on the NLDs as CTM, GHFB, and TGHFB are within the range of measured data. The MACS at $kT$=30 keV are 37.87 mb, 34.080 mb, and 34.46 mb {with} the NLDs as CTM, GHFB, and TGHFB, respectively. Our analysis of the MACS are within the range of the experimental results  $34\pm2.8$~mb \cite{Tessler},  $33\pm2$~mb \cite{Mutti}, and $37.3\pm4.3$~mb \cite{Walter} 32.133$\pm$2.36~mb~\cite{Tomyo}.}  {Furthermore, we have analyzed the MACS for the $^{86}$Kr(n,$\gamma$)$^{87}$Kr process. 
The Talys based calculated MACS along with the measured data for the $^{86}$Kr(n,$\gamma$)$^{87}$Kr radiative capture process are depicted in Fig.~(\ref{fig:2}).} {The present computed results of MACS based on the NLDs as CTM, BSFM, GSF, and SHFB agree with the measured results}. The MACS for $^{86}$Kr(n,$\gamma$)$^{87}$Kr  at $kT$=30 keV is 4.56 mb ({NLD as GHFB}) which {is} within the range of measured MACS, 3.5$\pm$03~mb~\cite{Kappeler9} and  3.8$\pm$07~mb~\cite{Kappeler10}.

Furthermore, we have computed the radiative capture rates based on best fitted NLD as CTM  and temperature dependent relativistic mean field model as radiative strength function for $n + {^{84}\rm{Kr}}\rightarrow {^{85}\rm{Kr}+{\gamma}}$ and $n + {^{86}\rm{Kr}}\rightarrow {^{87}\rm{Kr}+{\gamma}}$ processes up to $T_9$=1. The rates for the present analysis are mentioned in Table.~(\ref{tab:Rates}). It was found that the  radiative capture rates of $n + {^{84}\rm{Kr}}\rightarrow {^{85}\rm{Kr}+{\gamma}}$ {are} higher than the {rates of} $n + {^{86}\rm{Kr}}\rightarrow {^{87}\rm{Kr}+{\gamma}}$. The $^{86}\rm{Kr}$ is stable among the isotopes of Kr. It has neutron number $N$=50, which cause relatively smaller capture cross-section for the projectile neutrons at typical $s$-process temperature than $^{84}\rm{Kr}$ having $N$=48. Therefore, $^{84}\rm{Kr}$ does not offer any resistance to the incoming neutron for an interaction. It is the confirmation of the fact that the $^{86}$Kr(n,$\gamma$)$^{87}$Kr is a bottleneck of the $s$-process of nucleosynthesis.
\begin{table}[h!]
	\centering
	\caption{Comparision of the radiative capture rates. The first column shows the temperature in the units of $10^{9}$~K, the secound column show the radiative capture rates of $n + {^{84}\rm{Kr}}\rightarrow {^{85}\rm{Kr}+{\gamma}}$ and the third column show the radiative capture rates of  $n + {^{86}\rm{Kr}}\rightarrow {^{87}\rm{Kr}+{\gamma}}$.}
	\label{tab:Rates}       
	\addtolength{\tabcolsep}{1pt}
	\begin{tabular}{ccc}
		\toprule
		\multicolumn{1}{c}{} & \multicolumn{2}{c}{{Radiative capture rates ($cm^{3}mol^{-1}s^{-1}$)} }  \\
		\cmidrule(rl){2-3} 
		$T_9$ & $^{84}$Kr(n,$\gamma$)$^{85}$Kr& $^{86}$Kr(n,$\gamma$)$^{87}$Kr  \\
		\midrule
		0.0001 & 3.20$\times$10$^7$ & 3.80$\times$10$^6$ \\ 
		0.0005 & 1.99$\times$10$^7$ & 2.42$\times$10$^6$ \\ 
		0.001 & 1.53$\times$10$^7$ & 2.03$\times$10$^6$ \\ 
		0.005 & 9.98$\times$10$^6$& 1.79$\times$10$^6$ \\ 
		0.01 & 8.82$\times$10$^6$ & 1.63$\times$10$^6$ \\ 
		0.05 & 7.35$\times$10$^6$ & 1.16$\times$10$^6$ \\ 
		0.1 & 6.48$\times$10$^6$ & 9.77$\times$10$^5$ \\ 
		0.15 & 6.05$\times$10$^6$ & 8.97$\times$10$^5$ \\ 
		0.2 & 5.80$\times$10$^6$ & 8.53$\times$10$^5$ \\ 
		0.25 & 5.65$\times$10$^6$ & 8.26$\times$10$^5$ \\ 
		0.3 & 5.55$\times$10$^6$ & 8.08$\times$10$^5$ \\ 
		0.4 & 5.45$\times$10$^6$ & 7.87$\times$10$^5$ \\ 
		0.5 & 5.42$\times$10$^6$ & 7.75$\times$10$^5$ \\ 
		0.6 & 5.42$\times$10$^6$ & 7.69$\times$10$^5$ \\ 
		0.7 & 5.46$\times$10$^6$ & 7.67$\times$10$^5$ \\ 
		0.8 & 5.51$\times$10$^6$ & 7.67$\times$10$^5$ \\ 
		0.9 & 5.58$\times$10$^6$ & 7.70$\times$10$^5$ \\ 
		1.0& 5.67$\times$10$^6$ & 7.74$\times$10$^5$ \\ \hline
		\bottomrule
	\end{tabular}
\end{table}

\section{Conclusion}	
Within the framework of the {Talys v1.96 code}, the ($n$, $\gamma$) capture cross-section of the neutron magic number isotope $^{86}$Kr and nearly neutron magic number isotope $^{84}$Kr have been investigated over a wide energy spectrum. We have analyzed the capture cross-section via different level density models and temperature-dependent relativistic mean field model as {the radiative strength function}. Among the different level density models, the CTM is the best-fitted model for both the $^{84}$Kr(n,$\gamma$)$^{85}$Kr and {$^{86}$Kr(n,$\gamma$)$^{87}$Kr} processes. The respective stellar MACS at the standard $s$-process energy of $kT$=30 keV have shown a good comparison with the earlier reported results. For the formation of heavy elements in stars, we have analyzed the MACS and capture rates for both $^{84}$Kr(n,$\gamma$)$^{85}$Kr and $^{86}$Kr(n,$\gamma$)$^{87}$Kr processes. The present investigations shows that the computed rates of $^{84}$Kr(n,$\gamma$)$^{85}$Kr process are higher than the radiative capture rates of $^{86}$Kr(n,$\gamma$)$^{87}${Kr}. 
The fact that $^{86}$Kr has a magic number of neutrons, which adds stability and increases resistance to neutron capture, thereby reducing reaction rates, is consistent with the nuclear shell model predictions.
The found radiative capture rates confirm that from the studied reactions, the $^{86}$Kr(n,$\gamma$)$^{87}$Kr is the bottleneck of the $s$-process of nucleosynthesis.

\section*{Declarations}
\noindent \textbf{Conflict of Interest:} The authors declare no conflicts of interest regarding this article.


\newpage

\begin{thebibliography}{99}
\bibitem{Burbidge}	E.M. Burbidge, G.R. Burbidge, W.A. Fowler, and F. Hoyle, Rev. Mod. Phys. \textbf{29},   547 (1957)

\bibitem{Kappeler} F. Kappeler, Prog. Part. Nucl. Phys. \textbf{43}, 419  (1999) 

\bibitem{Pignatari} M. Pignatari, et al, Astrophys. J. \textbf{710}, 1557 (2010) 

\bibitem{Kappeler1} F. Kappeler, R. Gallino, S. Bisterzo, and W. Aoki, Rev. Mod. Phys. \textbf{83},  157 (2011)

\bibitem{Lugaro}	M. Lugaro, et al., Astrophys. J. \textbf{586, } 1305
  (2003)
\bibitem{Busso} M.Busso, R. Gallino, L.D. Lambert, C. Travaglio, and V.V. Smith,   Astrophys. J. \textbf{557},  802 (2001)
\bibitem{Tessler} M. Tessler et al Phys Rev C \textbf{104}, 015806  (2021) 

\bibitem{Mutti}	P. Mutti; H. Beer; A. Brusegan; F. Corvi; R. Gallino AIP Conference Proceedings \textbf{769}, 1327  (2005)

\bibitem{Walter} G. Walter, B. Leugers, F. Käppeler, Z. Y. Bao, G. Reffo \& F. Fabbri. NUCLEAR SCIENCE AND ENGINEERING: \textbf{93}, 357  (1986) 

\bibitem{Tomyo} A. Tomyo, Y. Nagai, T. Shima, H. Maki, K. Mishima, M. Segawaaand M. Igashira Nuclear Physics \textbf{A718}, 53O  (2003)  

\bibitem{Kappeler9} F. Kappeler, A. A. Naqvi and M. Al-Ohali Phys Rev C. \textbf{35}, 936  (1987)

\bibitem{TALYS1.96} A. J. Koning,  Hilaire, S., Goriely S. (2021). TALYS–1.96 A Nuclear Reaction Program, User Manual, Nuclear Research and Consultancy Group (NRG), Netherlands. http://www.talys.eu.	
\bibitem{hf} W. Hauser, H. Feshbach, Physical review, \textbf{87}, 366  (1952)
\bibitem{ompignore} D. Rochman, S. Goriely, A. J. Koning, H. Ferroukhi, Physics Letters B.
\textbf{764}, 109  (2017)
	\bibitem{localomp} A. J. Koning, J. P. Delaroche, Nuclear Physics A. \textbf{713}, 231  (2003)
	\bibitem{Kabir1} A. Kabir and  Jameel-Un Nabi, Nucl. Phys. A, \textbf{1007}, 122118 (2021).
	
	\bibitem{Kabir2} A. Kabir et al., Commun. Theor. Phys., \textbf{74}, 
	025301 (2022).
	
	\bibitem{Kabir3}  A. Kabir et al., Braz. J. Phys., \textbf{50 }, 112 (2020).
	
	\bibitem{Kabir4} A. Kabir et al., Astrophys. Space Sci., \textbf{365},  1 (2020).
\bibitem{koni} A.J. Koning, S. Hilaire, S. Goriely Nuclear Physics A \textbf{810}, 13  (2008)
\bibitem{shfb} S. Goriely, F. Tondeur, and J.M. Pearson. Atom. Data Nucl. Data Tables \textbf{77},  311  (2001)
\bibitem{ghfb} S. Goriely, S. Hilaire, and A.J. Koning. Phys. Rev. C \textbf{78},  064307  (2008)
\bibitem{tdghfb} S. Hilaire et al. Physical Review C \textbf{86},  064317 (2012)
\bibitem{rmf} I. Daoutidis and S. Goriely. Phys. Rev. C \textbf{86},  034328 (2012)
\bibitem{Kappeler10}G. Walter, H. Beer, F. Kappeler, and R. D. Penzhorn, Astron. Astrophys \textbf{155,} 247 (1986)
\bibitem{compiled}	For results compiled in evaluated nuclear data libraries International Atomic Energy Agency (IAEA) on
www-nds.iaea.org, or the OECD Nuclear Energy Agency on
www.nea.fr/html/dbdata/4.
	
\end{thebibliography}
\end{document}